\documentclass[mynatbib,surname,CE,MSC]{SYS-PV-JR}

\usepackage{setspace}

\usepackage{caption}

\newcommand\dsp{\displaystyle}
\newcommand\softwarename{SplitSup}

\newcommand*\patchAmsMathEnvironmentForLineno[1]{%
  \expandafter\let\csname old#1\expandafter\endcsname\csname #1\endcsname
  \expandafter\let\csname oldend#1\expandafter\endcsname\csname end#1\endcsname
  \renewenvironment{#1}%
     {\linenomath\csname old#1\endcsname}%
     {\csname oldend#1\endcsname\endlinenomath}}%
\newcommand*\patchBothAmsMathEnvironmentsForLineno[1]{%
  \patchAmsMathEnvironmentForLineno{#1}%
  \patchAmsMathEnvironmentForLineno{#1*}}%
\AtBeginDocument{%
\patchBothAmsMathEnvironmentsForLineno{equation}%
\patchBothAmsMathEnvironmentsForLineno{align}%
\patchBothAmsMathEnvironmentsForLineno{flalign}%
\patchBothAmsMathEnvironmentsForLineno{alignat}%
\patchBothAmsMathEnvironmentsForLineno{gather}%
\patchBothAmsMathEnvironmentsForLineno{multline}%
}

\volname{}

\jvolume{00}

\jshort{syv}

\jvol{}

\jissue{0}

\access{Advance Access publication Xxx XX, XXXX}

\cename{XX}

\pubyear{2016}

\copyyear{2016}

\artid{xxx}

\renewcommand\theequation{\arabic{equation}}
\begin{document}

\setcitestyle{authoryear,open={(},close={)}}

\title[Split scores and applications]{Split scores: a tool to quantify phylogenetic signal in genome-scale data}
\author[]{E\textsc{lizabeth S.} \surname{A\textsc{llman}}$^{1,\ast}$,
L\textsc{aura S.} \surname{K\textsc{ubatko}}$^{2}$,
J\textsc{ohn A.} \surname{R\textsc{hodes}}$^{1}$
}

\address{$^1$Department of Mathematics and Statistics, University of Alaska Fairbanks, Fairbanks, AK, USA
$^{2}$Department of Statistics, Department of Evolution, Ecology, and Organismal Biology, The Ohio State University, Columbus, OH, USA
$^{\ast}$Correspondence to be sent to: P.O. Box 756660, Fairbanks, AK, 99775-6660, USA \newline
E-mail: e.allman@alaska.edu\\
}

%

\abstract{ Detecting variation in the evolutionary process along
  chromosomes is increasingly important as whole-genome data become
  more widely available. For example, factors such as incomplete
  lineage sorting, horizontal gene transfer, and chromosomal inversion
  are expected to result in changes in the underlying gene trees along
  a chromosome, while changes in selective pressure and mutational
  rates for different genomic regions may lead to shifts in the
  underlying mutational process.  We propose the split score as a
  general method for quantifying support for a particular phylogenetic
  relationship within a genomic data set. Because the split score is
  based on algebraic properties of a matrix of site pattern
  frequencies, it can be rapidly computed, even for data sets that are
  large in the number of taxa and/or in the length of the alignment,
  providing an advantage over other methods (e.g., maximum likelihood)
  that are often used to assess such support. Using simulation we
  explore the properties of the split score, including its dependence
  on sequence length, branch length, size of a split and its ability
  to detect true splits in the underlying tree.  Using a sliding
  window analysis, we show that split scores can be used to detect
  changes in the underlying evolutionary process for genome-scale data
  from primates, mosquitoes, and viruses in a computationally
  efficient manner. Computation of the split score has been
  implemented in the software package SplitSup.
 [phylogenetic trees, split scores, genome-scale data analysis, general Markov model, matrix flattenings,
 singular value decomposition]}

\newpage

\maketitle

\nolinenumbers

Building on recent mathematical progress in understanding phylogenetic
models from an algebraic perspective, we develop here a new tool for
empirical analysis, the \emph{split score}. This score allows one to
compare support in a sequence alignment for different possible splits
(i.e., bipartitions of taxa, corresponding to putative edges in
trees). Through a ``sliding window'' analysis, we demonstrate how this
can be used to investigate a variety of biologically interesting
changes in evolutionary processes along the genome, such as differing
evolutionary trees, inversions, and changes in selective constraints.

Although similar analyses have been performed using full maximum
likelihood inference of trees \citep{Hobolth2007,Boussau2009}, the
split score offers several advantages over previous methods: (1) it
focuses directly and solely on a split, and not on the split as
inferred with a full tree and model parameters, (2) it is
theoretically justified for models of sequence evolution beyond those
routinely assumed, in particular requiring neither a stationary
distribution, nor homogeneity of the substitution process over the
tree, and (3) its computation is extremely fast, even for a large
number of taxa, making it a viable tool for exploratory
analyses. While some caution is necessary in interpreting and
comparing split scores, our empirical examples show they can provide
biological insight.

\smallskip

The thread of ideas we build on to develop the split score is perhaps
not widely known to empiricists, though the field of phylogenetics has
benefitted in numerous ways from the application of ideas from algebra
and geometry.  One of the earliest algebraically-based methods for
inferring an evolutionary tree was ``evolutionary parsimony,'' put
forth by \citet{lake1987}, in which simple weighted sums of estimated
site pattern probabilities were used to infer phylogenetic
relationships. Viewing these sums as linear (i.e., first degree)
polynomials, they are an instance of {\itshape phylogenetic
  invariants} --- polynomials whose values should be zero when
evaluated at the site pattern probabilities for a particular tree and
substitution model.  Independently, \citet{cavenderfelsenstein1987}
proposed higher degree polynomial invariants as a tool for inference,
but their detailed work was with a 2-state model, and not developed
for practical application.
 
Following these initial works, the ideas were extended in various ways
\citep{Cavender1989, Fu1992, EvansSpeed93, SteSzErWa93, Fu1995,
  Hendy1996, ARmb}. However, the use of invariants in empirical
phylogenetic studies was rare. Simulations showed Lake's linear
invariants needed significantly longer data sequences to produce
accurate estimates than traditional methods such as maximum likelihood
\citep{hillisetal1994,huelsenbeck1995}, while understanding of higher
degree invariants was still insufficient for their practical
use. Despite this unpromising start, it is interesting to note that
similar invariant methods are currently the basis of widely-used
analysis tools, though this connection is rarely mentioned in the
current literature.  For example, the ABBA-BABA test
\citep{durandetal2011}, used to detect introgression and hybrid
speciation in empirical data, is based on the difference in the
empirical frequencies of two types of site patterns among four taxa
(ABBA-like patterns and BABA-like patterns).

Indeed, the use of an algebraic framework for phylogenetic inference
and theory advancement gained traction only after mathematical
understanding of higher-degree polynomial relationships in site
pattern probabilities was further developed.  Using ideas primarily
from algebraic geometry, large classes of informative phylogenetic
invariants for a variety of models were identified and
characterized. Often these non-linear invariants can be linked to
local features in a tree, such as a vertex or edge \cite{SStoric2005,
  ARgm, RSlargemixtures2012}.

Such advances led to the development of arrangements of the
probability distribution of site patterns on large trees into arrays
of reduced dimensions, by ``flattening'' the distribution according to
edges or nodes within a phylogenetic tree.  For example, an edge
flattening is a matrix whose rows correspond to possible site patterns
for the taxa on one side of the edge, and columns to possible site
patterns on the other. An entry of the matrix is the probability of
observing the amalgamated site pattern for its row and column.
Flattenings were first presented broadly to the research community in
the 2004 arXiv preprint of \citet{ARgm}, to understand the ideal of
all invariants for the general Markov (GM) model on trees.

Importantly, the study of algebraic characteristics of such
flattenings led to the development both of methods for establishing
identifiability of gene tree topologies and associated parameters, and
to the development of algorithms for inferring the tree.  For example,
rank conditions on matrix flattenings were used by \citet{ARidtree} to
identify tree topologies for $k$-class mixtures ($k =1, 2, 3$) of GM
models and independently by \citet{eriksson2005} for the non-mixture
GM model.  The latter work also made the first use of the singular
value decomposition (SVD) of matrix edge flattenings, as a tool for
measuring approximate matrix rank, to develop an invariant-based
algorithm for tree construction.  Although performance of that early
algorithm was disappointing, several recent works have explored the
use of the SVD of flattenings for tree inference in ways that appear
much more promising
\citep{casanellasfernandezsanchez2007,casanellasfernandezsanchez2015}.

These algebraically-motivated ideas have also been applied to the
multi-locus setting, in which estimation of a species trees under the
coalescent model is the goal.  Again using the ideas of flattenings
and rank approximations, \citet{chifmankubatko2015} derived invariants
for the $4$-taxon species trees under the coalescent model based on
the site pattern probability distribution, leading to both
establishment of identifiability of the species tree, and an inference
method called SVDQuartets \citep{chifmankubatko2014} that is
implemented in PAUP* \citep{swofford2016}. Other work in this area
includes invariant-based methods of establishing identifiability of
the species tree from collections of gene tree topologies
\cite{allmanetal2011a} or from clade probabilities
\cite{allmanetal2011b}.

\ 

Rather than consider inference of an entire phylogeny, here we turn
our attention to the use of tools arising from an algebraic
phylogenetic framework to learn about various features of large-scale
genomic data.  In particular, we consider the case of data arising
from a single gene phylogeny and show how a statistic based on the
singular value decomposition can be used to measure support for
particular phylogenetic relationships in that data.  We study the
behavior of this statistic using simulated data to demonstrate the
impact of factors such as the length of the gene, the branch lengths
in the true underlying gene tree, and the substitution model.

We then demonstrate how our statistic can be applied to whole-genome
data to extract features of the underlying evolutionary model, both
with regard to the tree structure and with regard to the substitution
process, by applying our method to three empirical data sets. The
first is the data of \citet{pattersonetal2006}, which consists of
whole-genome data for five primate species.  These data demonstrate
the ability of our method to detect the gene-level variability
predicted by the coalescent process.  The second example uses
whole-genome data for a species complex of {\itshape Anopheles
  gambiae} mosquitoes from \citet{fontaineetal2015}, for which our
method is able to detect the region of a known chromosomal
inversion. Finally, we apply our method to genome-scale data from 29
whole-genomes of Cassava Brown Streak Virus and Ugandan Cassava Brown
Streak Virus, demonstrating that the method captures variation in the
substitution process from gene-to-gene across the viral genome.  These
examples highlight that a major advantage of our method is the rapid
computation time, with an entire chromosome being analyzed in a matter
of minutes.

We begin by providing the mathematical theory underlying our proposed
method. Readers interested primarily in the application of the
methodology can skip to the Methods section of the paper, where we
present our statistic and describe how it can be used to analyze
large-scale empirical data.

\section*{Theoretical Background}

\subsection*{Basic Theory}
The \emph{general Markov} (GM) model of evolution of DNA sequences on
trees underlies the theoretical development of our statistic, the
split score. This model assumes an arbitrary probability distribution,
$\boldsymbol\pi=(\pi_A,\pi_G,\pi_C,\pi_T),$ describing bases at the
root of the tree. In addition, to each edge, $e$, of the tree
(directed away from the root) is associated a $4\times4$ matrix,
$M_e$, of conditional probabilities of the various base
substitutions. No special relationships between the matrices
associated to different edges is assumed; in particular, the GM model
does not assume time-reversibility of the substitution process, a
stationary base distribution, homogeneity of the substitution
processes across the edges of the tree, nor even the existence of an
underlying homogeneous continuous-time process on any edge. This model
thus encompasses, but is more general than, the general
time-reversible (GTR) model and its submodels which are commonly used
in current data analysis. However, it lacks the across-site rate
variation features that are also often combined with the GTR model in
the form of invariant sites and $\Gamma$-distributed scaling factors
(e.g., GTR+I+$\Gamma$).

The GM model implies that certain \emph{conditional independence
  statements} hold for the joint distribution of bases at the leaves
of the tree. These express the fact that the base substitutions that
occur in a clade on a rooted tree are not affected by those occurring
outside the clade, except through the sequences at the clade's most
recent common ancestor.  To be more precise, pick any edge $e$ of the
unrooted tree and let $v$ be one of its end nodes. Deleting $e$ from
the tree breaks it into two parts, and induces a partition of the taxa
$X$ into disjoint sets $X_1$ and $X_2$, the \emph{split} $X_1|X_2$
associated to $e$. Then the joint distribution of bases at the leaves
of the tree can be organized as a $4^{|X_1|}\times 4^{|X_2|}$ matrix
$F_e$, with rows indexed by patterns of bases for $X_1$, and columns
by patterns of bases for $X_2$. This is the \emph{edge flattening} of
the joint distribution along $e$. The conditional independence
statement above is then formulated mathematically as the fact that
$F_e$ has a factorization
\begin{linenomath*}
$$
F_e=M_1^T D M_2
$$
\end{linenomath*}
where $D$ is a $4\times4$ diagonal matrix with entries giving the base
distribution at $v$, and the $M_i$ are $4\times 4^{|X_i|}$ stochastic
matrices giving probabilities of the bases at the taxa in $X_i$
conditioned on the bases at $v$. As a consequence of the matrix
factorization, the rank of the matrix $F_e$ will be at most 4.

One can as well consider \emph{any} split $X_1|X_2$ of the taxa,
whether associated to an edge or not, and then construct a split
flattening $F$ of the joint distribution according to it. If the split
does not arise from an edge of the tree, then $F$ does not have the
simple structure above. Under very mild and plausible assumptions on
the nature of the model parameters, this implies the rank of the
matrix $F$ is larger than 4 \cite{ARidtree, eriksson2005}. 

The central idea of our method is to view an empirical
distribution of bases in data sequences as an approximation of the
true distribution, and then use a measure of how close a split
flattening of this empirical distribution is to a matrix of rank 4 as
an indication of whether the split is supported or not. If we have
exact distributions arising from the general Markov model, our measure
will be zero for splits displayed on the tree, and positive for splits not displayed.

\subsection*{Rank-4 matrix approximations, the SVD, and split scores}

One way of measuring the size of a matrix uses the \emph{Frobenius
  norm}; if $M=(m_{ij})$, then
\begin{linenomath*}
$$
||M||=||M||_F=\left (\sum_{i,j} m_{ij}^2\right )^{1/2}.
$$
\end{linenomath*} 
The associated distance between two matrices $F$ and $G$ is then
$||F-G||$. Adopting these measures means there is a good tool to
determine the closest rank-4 matrix to a given matrix, using the
\emph{singular value decomposition} (SVD) and software developed for
computing it.

The SVD of an $m\times n$ real matrix $F$ is a factorization
\begin{linenomath*}
$$
F=U\Sigma V^T,
$$
\end{linenomath*}
where $U$ and $V$ are $m\times m$ and $n\times n$ orthogonal matrices,
and $\Sigma$ is a $m\times n$ diagonal matrix with
entries 
\begin{linenomath*}
$$
\sigma_1\ge \sigma_2\ge \cdots \ge \sigma_{\min(m,n)}\ge
0,
$$
\end{linenomath*} 
the singular values of $F$.  By the Eckart-Young Theorem
\cite{EY36}, under the Frobenius norm the closest rank-4 approximation
to a matrix $F$ is $\tilde F =U\tilde \Sigma V^T$ where $\tilde
\Sigma$ is obtained from $\Sigma$ by zeroing out all but the 4 largest
singular values. Moreover the Frobenius distance between $F$ and
$\tilde F$ is
\begin{linenomath*}
$$
||F-\tilde F||=\left( \sum_{i=5}^{\min(m,n)} \sigma_i^2\right) ^{1/2}.
$$
\end{linenomath*}
As a measure of split support, then, we define the \emph{split score}
\begin{linenomath*}
\begin{multline*}
S(X_1|X_2)=S(F)= \left (\frac{ \sum_{i=5}^{\min(m,n)}
    \sigma_i^2}{\sum_{i=1}^{\min(m,n)} \sigma_i^2} \right)^{1/2}\
    =\\ \ \left(1-\frac{ \sum_{i=1}^{4} \sigma_i^2}{\sum_{i=1}^{\min(m,n)}   \sigma_i^2} \right)^{1/2},
\end{multline*}
\end{linenomath*}
where $F$ is the $(X_1|X_2)$-flattening of the empirical distribution.
The denominator is introduced so that the result is independent of the
scaling of $F$. Thus the formula may be applied either to $F$, or to
the unnormalized matrix of counts leading to it.  The split score
takes values in the interval from 0 to 1. A score of 0 indicates $F$
is a rank 4 matrix, and a positive score indicates that it is not.
Implicit in this theory is that split scores are defined for
``gapless'' alignments.  We exclude any site at which one or more of
the taxa has a gap, or missing data of any kind.

To compute split scores, one must compute singular values of
potentially large matrices. Since the Frobenius norm is related to the
singular values by
\begin{linenomath*}
$$
||F||=\left( \sum_{i=1}^{\min(m,n)} \sigma_i^2\right)^{1/2},
$$
\end{linenomath*}
the formula above can also be written as
\begin{linenomath*}
$$
S(X_1|X_2)=S(F)=\left (1-\frac{ \sum_{i=1}^{4} \sigma_i^2}{||F||^2 }\right)^{1/2}.
$$
\end{linenomath*}
Using this formula, only the 4 largest singular values are needed.
This observation provides significant computational advantage, as there are good
algorithms for computing a specified number of the largest singular
values with significantly faster runtimes than if all are needed.

If the number of taxa is large, and sequence lengths are as typical in
alignable sequences, any split flattening $F$ will be a large and
sparse matrix, i.e., most entries will be zero. Computation of
singular values of such matrices requires a sparse encoding of them in
software, and special packages for the SVD computation. Fortunately,
these are highly developed as the SVD has many applications in
scientific computing.  

Although one might suppose that computations would be slowed
considerably by increasing the number of taxa, thus exponentially
increasing the size of the flattening matrix, this is not the
case. Since site patterns appearing in finite-length sequence data
tend to be those more strongly reflecting the underlying tree, the
sparsity of the matrix tends to be patterned, with many zero rows or
columns which can be ignored.  Even though increasing sequence length
does lead to more non-zero entries in the matrix, this happens slowly
due to the very low probability of many site patterns. Finally, the
iterative routines used for computing singular values converge most
quickly to determine the few largest values, which are precisely the
ones we need.  Any detailed analysis of theoretical running time is
complicated by how the matrix sparsity and the size of the singular
values depend upon the phylogenetic model parameters, number of taxa,
and sequence length; still, one should expect fast performance.

In practice, we have found that assembling the sparse flattening
matrix dominated the computation time, since each sparse encoding
requires a scan of all unique site patterns in the alignment.
Nonetheless, all the necessary computations to produce split scores
for reasonable size data sets can be performed quickly enough that
runtime is of little concern.  For instance, for simulated data on a
100-taxon tree (not shown) with sequence length 1000 bp (respectively
10,000 bp) computing all 97 scores for splits displayed on the tree
took 0.186 seconds (respectively 8.68 seconds) on a MacBook Pro 3.1
GHz with 16 GB of memory. Computation time was similar for scores of
97 random splits of the same sizes.  See also the applications section
for an example of timing on empirical data.

\subsection*{Interpretation of split scores}

When applied to an empirical joint distribution, a low split score
(close to 0) indicates support for that split, and a higher score
(close to 1) indicates lack of support. However, a variety of
factors, such as split size, edge lengths, sequence length, and model
fit affect the interpretation of $S(X_1|X_2)$. Some of these effects
will be illustrated through the simulations described below.  Here we  
focus attention on one theoretical principle concerning the size of the split 
and its influence on the score.

Some of the effect of the size of a split (the number of taxa included
in each of the two groups) on the split score has a clear mathematical
explanation. The space of $m\times n$ matrices has dimension $mn$,
while the subset of those that have rank 4 or less forms an object of
dimension $4(m+n)-16$. 
(The simplest way to see this is to count free parameters
in the LU matrix factorization of an $m \times n$ matrix of rank 4.)

Applied to $F$, for $|X|=N$, $|X_i|=N_i$, we
have $m=4^{N_1}$, $n=4^{N_2}$, so
\begin{linenomath*}
$$
mn=4^N,\ \ 4(m+n)-16=4(4^{N_1}+4^{N-N_1})-16.
$$
\end{linenomath*} 
This last expression is easily seen to decrease as $N_1$ goes from 1
to $\lfloor \frac{N}{2} \rfloor$, where it has a minimum. That is, the
dimension of the set of matrices of rank at most 4 drops with the size
of $N_1$ until $N_1=N_2$ if $N$ is even, or $N_1 = N_2 - 1$ if $N$ is
odd. The smaller the dimension of the set of rank 4 matrices is, the
greater the distance should be between this set and a random
perturbation of one of its elements.  (To see this, imagine moving a
point in $3$-space that lies on a line $\ell$ contained in a plane
$\mathcal P$ a fixed distance in a random direction.  The movement
typically leaves the point further from the line $\ell$ than from
$\mathcal P$, since more of the motion will be in a direction within
the plane than in the direction of the line.)  Thus even for splits
arising from the tree on which data was simulated, we should expect
larger split scores for splits that are closer to balanced.  Indeed,
this geometric understanding explains why the tree reconstruction
algorithm of \citet{eriksson2005} performs poorly in practice, tending
to create trees with a preponderance of cherries and small clades.  By
comparing splits of different sizes, splits with size $N_1 = 2$ or
more generally $N_1 = k$ where $k$ is small are preferred and bias the
reconstruction.

We note that this dimensional understanding can be developed into a
theoretical correction to the split score which, at least
asymptotically, can overcome the dependence on split size. However, we
found this correction inadequate to substantially improve
comparability of the scores, so do not present it here.

\section*{Methods}

As described above, our method involves computation of a split score
associated with a putative edge $e$ of a phylogenetic tree.  If $X$
denotes the set of taxon names, then a \emph{split} $X_1 \mid X_2$ of
the taxa is a bipartition of $X$ into two disjoint sets $X_1$, $X_2$.
When $k = \vert X_1 \vert \le \vert X_2 \vert$ we call such a split a
\emph{$k$-split}.

For a $N$-taxon unrooted binary tree $T$, there are $2N-3$ \emph{true
  splits} on $T$, corresponding to edges of $T$. Of these, $N$ are
trivial splits (the $1$-splits which appear on all $T$), which we no
longer consider.  If $T$ is the true tree describing the evolutionary
history of the taxa under study, then all other splits are \emph{false
  splits}, since they do not correspond to edges in $T$.

Suppose now that $e$ is a putative edge in the true tree relating data
sequences and that $e$ corresponds to the split $X_1 \mid X_2$.  We
can arrange the observed counts of site patterns from the alignment of
a single gene into a $4^{|X_1|} \times 4^{|X_2|}$ matrix, called a
{\itshape flattening} and denoted by $F_e$ as above, where the rows of
$F_e$ are indexed by possible nucleotides for each of the taxa in
$X_1$ and the columns are indexed by possible nucleotides for each of
the taxa in $X_2$.  We assess support for the specified edge as a true
edge in the underlying phylogenetic tree by measuring how close the
matrix $F_e$ is to the nearest rank 4 matrix, based on the theory
described above.  We measure closeness using the split score
\begin{linenomath*}
\begin{equation}
S(X_1|X_2)=\left (1-\frac{ \sum_{i=1}^{4} \hat{\sigma}_i^2}{||F||^2 }\right)^{1/2}
\end{equation}
\end{linenomath*}
where $\hat{\sigma_i}$ refers to the $i^{th}$ singular value obtained
from $F_e$ and $||F||$ is as defined above.  

We have implemented the computation of split scores in the program
{\tt \softwarename} written jointly by the authors in the C
programming language using the publicly available SVDLIBC library.
This code, available at {\tt https://github.com/eallman/SplitSup/},
reads an alignment in PHYLIP format and returns either (i) a set of
scores corresponding to a list of user-provided splits; or (ii) the
values of a split score in a sliding-window across the length of the
alignment.  For option (ii), the user specifies the window size, the
number of nucleotides to move the window for the next computation, and
the minimum number of sites without gaps required to compute scores in
each window.  The SVD computations have been sped up significantly by
using a binary encoding of site patterns, and a sparse encoding of the
flattening matrices.

\smallskip

We next describe our methods for assessing the utility of the split score using both simulated and
empirical data. 
Although all simulations described below were carried out using the Jukes-Cantor model, 
we note that similar results can be obtained under any of the commonly used substitution models 
that are submodels of the GTR model, or more generally any submodel of the GM model. Indeed, 
this generality is one of the key features of the split score.
%
\begin{figure}[h]
\centerline{{\vbox to 175pt{\vfill\hbox to 250pt{\hfill\mbox{\includegraphics[width=.5\textwidth]{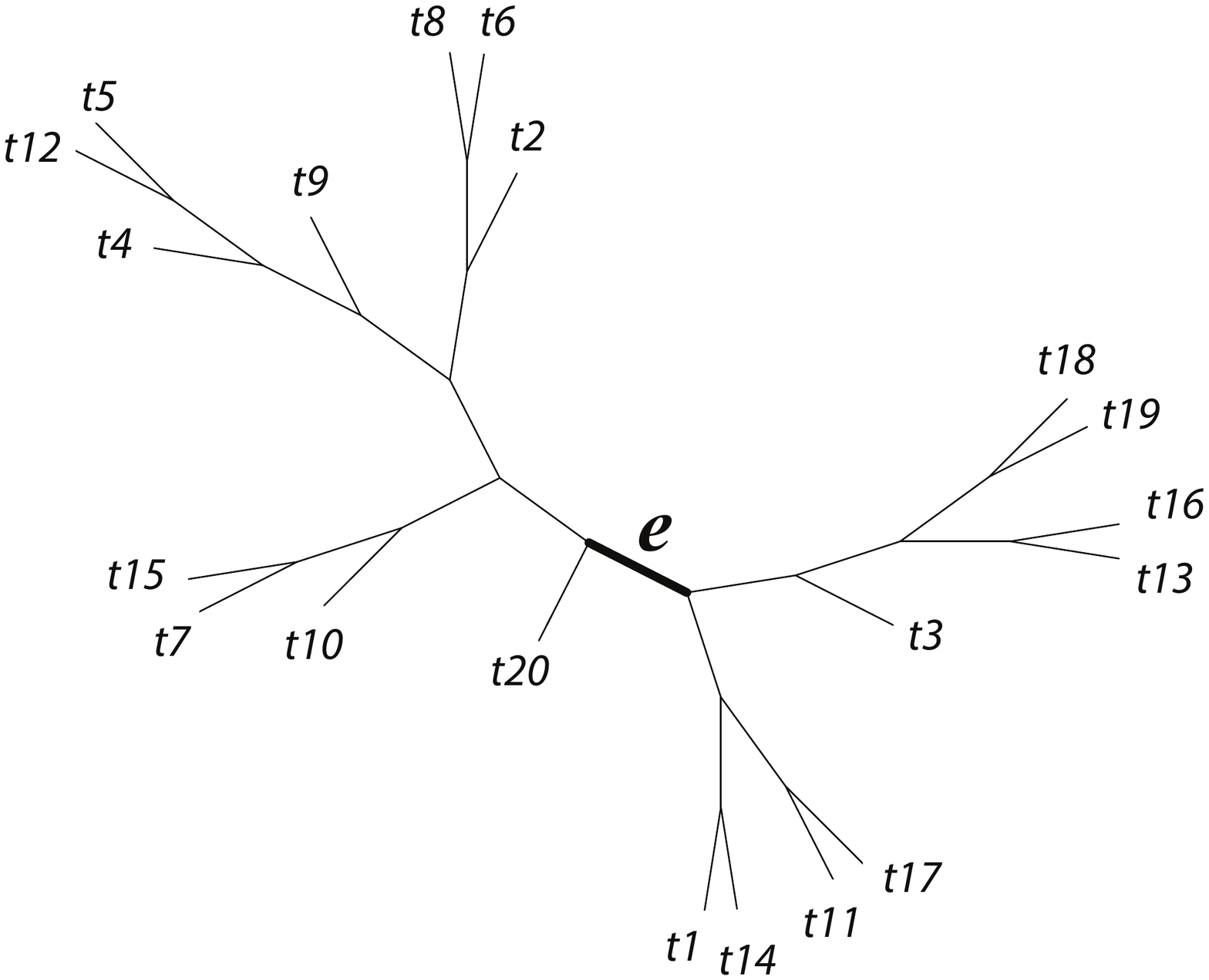}}\hfill}\vfill}}}
\caption{20-taxon tree used for simulations.}\label{fig:modeltree20tax}
\end{figure}

\subsection*{Simulation Study 1}

Our first simulations were designed to test that our split score can
detect true splits in the underlying tree, under ideal circumstances.
We also investigated how the magnitude and spread of splits scores
varies, and if there are qualitative differences between true and
false splits.  To accomplish this, we used SeqGen
\citep{rambautgrassly1997} to generate a single data set under the
Jukes-Cantor model for the tree in Figure \ref{fig:modeltree20tax}
with all branch lengths set to $0.05$.  For each value $k = 2, 3,
\dots, 10$, we computed the split score for each of the
$\binom{20}{k}$ possible splits and then generated histograms for the
scores for $k$-splits.  The $N-3$ non-trivial true $k$-splits were
marked on the appropriate histograms.

This enabled us to compare the score values for true and false splits,
understand the distribution of split scores, and gain insight into the
effect of \emph{split size} on the score; that is, how does proximity
of an edge to the tips of the tree or nearer the middle of the tree
affect the score's value?  To refine our understanding further, we
classified false splits by how ``far'' they were from a true split.
For example, if the true split of size $9$ in a tree consisted of $X_1
= $ \{taxa 1-9\} and $X_2 =$ \{taxa 10-20\}, then we say that a split
with $X_1 =$ \{taxa 1-8, taxon 10\} and $X_2 =$ \{taxon 9, taxa
11-20\} is one swap away from a true split.

\subsection*{Simulation Study 2}
Our next simulations were performed to explore the effects of sequence
length and branch length on the distribution of scores.  On the
$20$-taxon tree shown in Figure \ref{fig:modeltree20tax} we used
SeqGen to simulate 100 data sets of varying sequence lengths (500 bp;
5,000 bp; and 50,000 bp) with all branch lengths set to $0.05$ under
the Jukes-Cantor model.  To test the effect of tree diameter, we
simulated 100 replicate data sets of 500 bp under the Jukes-Cantor
model with all branch lengths set to $0.0125$, $0.025$, $0.05$.  We
then compared empirical distributions of split scores for all true
splits.
 
To improve our understanding of the effect of metric depth on the
distribution of a particular split score, 100 replicate data sets of
length 500 bp were next simulated under the Jukes-Cantor model with
various scalings.  Attention was focused on the $9$-split induced by
edge $e$ pictured in Figure \ref{fig:modeltree20tax}.  Scores for this
true split were computed on data simulated when all the branch lengths
were scaled by a fixed factor, every branch was scaled except $e$,
only edge $e$ was scaled, and when all edges to one side of the split
were scaled by the factor.  In short, with either a subset or all of
the branches rescaled, split scores were compared.]

A last simulation was a hybrid of the previous two.  Here we focused
attention on a true $9$-split and a true $2$-split in the tree, and
carefully selected other bipartitions of the 20 taxa that we
considered `close' ($1$-swap or $2$-swap) or `distant' (random).  This
made for a total of seven splits (four $9$-splits, three $2$ splits),
\{t9, s9-1, s9-2, s9-R, t2, s2-1, s2-R\}, which we now describe.

The true $9$-split, t9, is the one pictured in Figure
\ref{fig:modeltree20tax}.  By interchanging taxa $3$ and $20$ we
obtain split s9-1 (for \emph{nine} split, \emph{one} swap away from
t9).  By interchanging taxa \{1, 3\} and taxa \{10, 20\}, we obtain
split s9-2 (for \emph{nine} split, a \emph{two} swap away from t9).
Because the swaps here interchange only a few ``topologically close''
taxa, these false splits are expected to be difficult to distinguish
from true splits.  The split s9-R (for \emph{Random nine} split) has
taxa \{2 3 6 7 14 15 16 18 19\} on one side of the bipartition, and
should be easier to distinguish from a true one.

The split t2 is the true $2$-split grouping together taxa \{1 14\}.
The split s2-1 groups together taxa \{1 11\}, resulting from a 1-swap
of ``topologically close'' taxa, and is expected to be hard to
distinguish as false. The split s2-R groups taxa \{1 8\} together, and
since these are far from one another in the tree, should be easier to
distinguish as false.

For the simulations, 100 replicate data sets were made under the
Jukes-Cantor model on the tree (Fig \ref{fig:modeltree20tax}) with
branch lengths set to 0.05, and sequence lengths of 500 bp, 5000 bp,
and 50000 bp.  Additionally, fixing the sequence length at 500 bp, 100
replicate data sets were simulated when the branch lengths were set to
0.0125, 0.025, and 0.05.  For each parameter setting and each
replicate, splits scores were computed for the seven splits.

To test that the true split scores were the lowest, or closest to the
lowest, we computed the difference between the scores of the false
splits and the true split (e.g., compute score(s9-1) - score(t9)).  A
positive difference indicates that the true score is the smallest and
the magnitude of the difference reveals how close in value are the
scores of nearby and distant splits under a variety of model settings.

\subsection*{Simulation Study 3}
\begin{figure*}[!ht]
\centerline{{\vbox to 300pt{\vfill\hbox to 500pt{\hfill\mbox{\includegraphics[width=\textwidth]{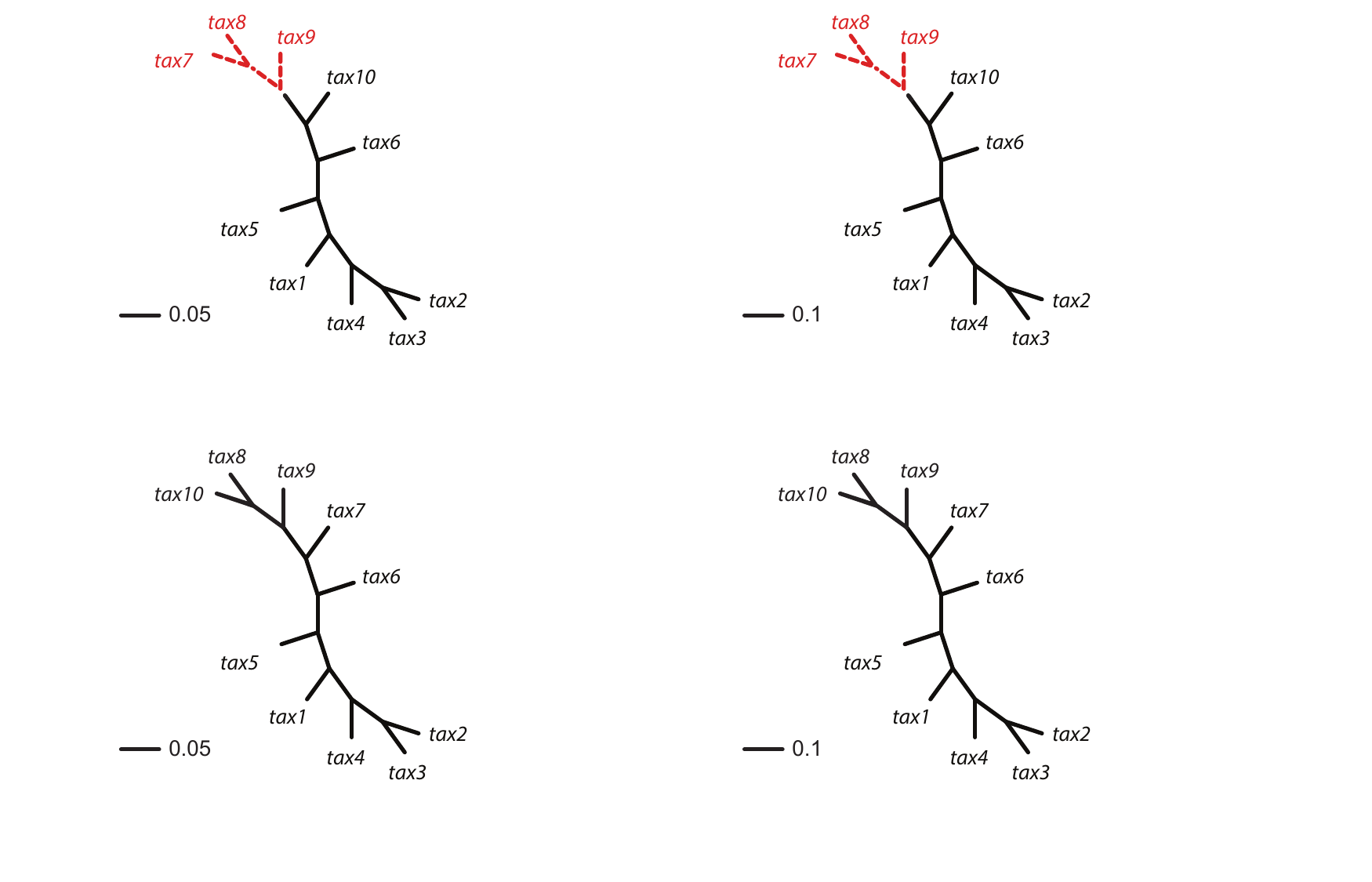}}\hfill}\vfill}}}
\caption{10-taxon trees used for sliding window simulations in Simulation Study 3.
(See online for color.)}\label{fig:slidingwindowsim}
\end{figure*}
We evaluated our method on genome-scale data using a sliding-window
computation of split scores for contiguous sections along the
alignment.  To test this approach, we used Seq-gen and the
Jukes-Cantor model to simulate data for an alignment of total length
20,000 bp, by concatenating four chunks of 5,000 bp simulated on the
$10$-taxon model trees shown in Figure \ref{fig:slidingwindowsim}.
The first two trees are topologically identical, but the branch
lengths of $0.05$ in tree 1 are doubled to a value of $0.1$ in tree
$2$.  The second two trees are also topologically identical with all
branch lengths $0.05$ and $0.1$ respectively; however, these trees are
topologically distinct from trees 1 and 2 in that taxa $7$ and $10$ in
tree 1 have been interchanged to obtain the topology of trees 3 and 4.
In particular taxa $\{7, 8, 9\}$ form a true $3$-split in trees 1 and
2, and taxa $\{8, 9, 10\}$ form a true $3$-split in trees 3 and 4.
The sliding window analysis computed and compared scores for the
$3$-splits $\{7,8,9\}$ and $\{8,9,10\}$ using a window of size $500$
bp at at intervals of 100 bp (slide offset size).  This means the
start sites for each 500 bp window were $1, 101, 201,$ etc.

\subsection*{Application to Empirical Data}

Our simulations demonstrate that the split score can be used to detect
both changes in topological relationships and changes in the
underlying evolutionary process. We now demonstrate the utility of the
split score in practice by application to three empirical data sets.
Note first that alignments of empirical data often have sites in which
some sequences have gaps, which are excluded for the computation of
the split score. If gapped sites are too numerous this can result in
little data being left to analyze, and split scores can be
misleading. It is therefore wise to require a minimum number of
non-gapped sites to eliminate spurious signals. We specify the minimum
number of sites required in each of analyses below.

\smallskip

\noindent {\bf Primate Data}-- We applied the sliding window method to
the data set of \citet{pattersonetal2006}, which consists of
whole-genome data for human, chimp, gorilla, orangutan, and macaque.
(Data at \url{http://genetics.med.harvard.edu/reich/Reich_Lab/} \url{Datasets_-_Patterson_2006.html})
We considered the data for chromosome 7 ($\sim$ 1.9-million bp), and
applied our method with a window size of $10,000$ bp, a slide size of
$1,000$ bp, and required that at least $500$ (= 5\%) non-gap sites
were present in a window for a score to be computed.  We computed
scores for the three possible splits of the taxa human, chimp,
gorilla, and orangutan.  For each window examined, we determined which
of the three splits gave the lowest split score, and we plotted the
results using different colors/shading to indicate which split each
region of the genome supported most strongly.  The primary process
leading to variation in the genealogy across a chromosome for these
taxa is expected to be the process of incomplete lineage
sorting. Thus, we expect that the majority of the data will support
the human-chimp $|$ gorilla-orangutan split most strongly, with
approximately equal support for the other two splits.

\smallskip

\noindent {\bf Mosquito Data} -- \citet{fontaineetal2015} carried out
a phylogenomic analysis of whole genomes from the 
{\itshape Anopheles  gambiae} species complex.  
(Data at \url{http://dx.doi.org/10.5061/dryad.tn47c}.)
We considered the analysis of chromosome
2L, and utilized an approximately 37.5-million bp alignment of a
subset of this chromosome for four species: {\itshape An. gambiae},
{\itshape An. coluzzii}, {\itshape An. arabiensis}, and 
{\itshape  An. christyi} (the outgroup).  As with the primate data, we
considered all three possible splits, and carried out the sliding
window analysis with a window size of $10,000$ bp, a slide size of
$1,000$ bp, and required that at least 500 (= 5\%) non-gap sites were
present in a window for a score to be
computed. \citet{fontaineetal2015} found that gene flow between the
ancestor of the {\itshape An. gambiae}-{\itshape An. coluzzii} clade
occurred with {\itshape An. arabiensis}, revealing an interesting
pattern in the region of a known chromosomal inversion on chromosome
2L.  In particular, because both {\itshape An. coluzzii} and 
{\itshape  An. arabiensis} experienced the inversion, while 
{\itshape  An. gambiae} did not, the tree supporting a sister relationship
between {\itshape An. coluzzii} and {\itshape An. arabiensis} is
expected to dominate in this region, while {\itshape An. gambiae} and
{\itshape An. coluzzii} are expected to be sister taxa elsewhere along
the chromosome.  We thus assess whether our analysis can detect the
region of this chromosomal inversion.

\smallskip

\noindent {\bf CBSV Data} -- \citet{alicaietal2016} recently collected
complete viral genomes for 14 samples of Cassava Brown Streak Virus
(CBSV) and 15 samples of Ugandan Cassava Brown Streak Virus (UCBSV).
The viral genomes consist of 10 distinct genes, and we
considered sequence data for the entire genome (i.e., all 10 genes)
for all 29 individual samples.  (See Figure 3 of Alicai et
al. (2016)\nocite{alicaietal2016} for Genbank accession numbers.)
The published phylogenetic analysis based on these data indicate that
CBSV has an accelerated rate of evolution compared to UCBSV, which
matches field observations indicating increased virulence for these
strains.  We applied our method with a window size of $500$ bp, a
slide size of $100$ bp, and required that at least 100 non-gap sites
were present in a window for a score to be computed. We considered the
single split that partitioned the sequences into CBSV versus UCBSV,
and evaluated changes in the score across the genome as an indicator
of which genes may be involved in the shift in evolutionary rate of
CBSV.  This example highlights application of our method to a data set
of more than four taxa when gene boundaries are known.

\section*{Results}

\subsection*{Simulation 1 Results}

\smallskip

\noindent \textbf{Identifying true splits:}  
Displayed in Figure \ref{fig:splitscoredist} are split scores
distributions of all $k$-splits for $k = 2, 4, 6, 9$. (Histograms for 
other values of $k$ are not shown but fit the pattern seen here).
The values of our scores for true splits in the tree are shown with red dots.
For all $k = 2, 3, \dots, 10$, scores for true splits from the generating
tree are the \emph{smallest} in the distributions.  This shows that even
for simulated data of modest size (500 bp) the split score picks out
true splits in the tree.
%
%
\begin{linenomath*}
\begin{figure*}[!htb]
\centerline{\vbox to 375pt{\vfill\hbox to 800pt{\hfill\mbox{\includegraphics[width=\textwidth]{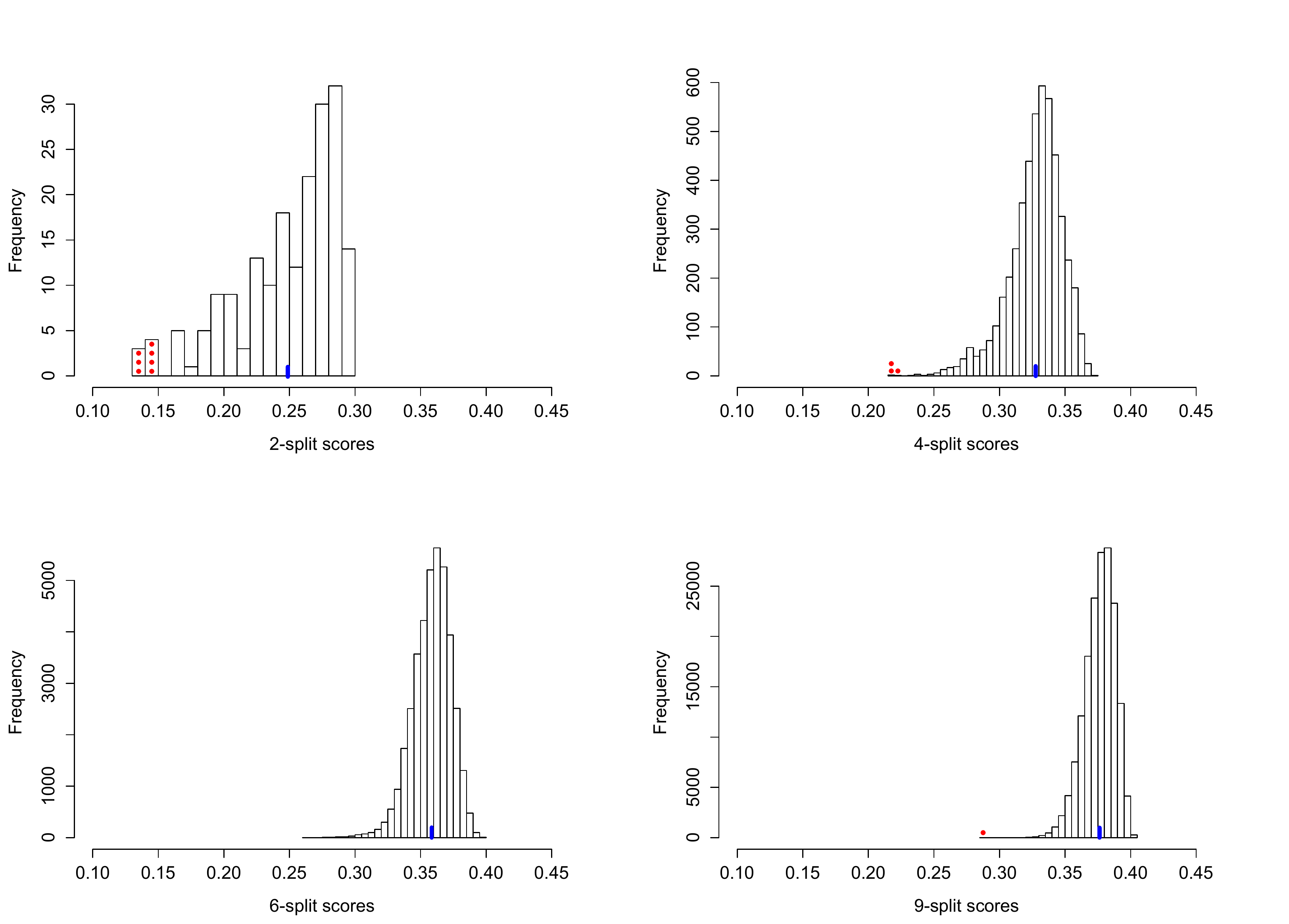}}\hfill}\vfill}}
\caption{The distributions of $k$-split scores for $k = 2, 4, 6, 9$ are shown.  Scores were computed
for all $\binom{20}{k}$ possible splits for a data set of 500 bp generated under the
Jukes-Cantor model on the tree in Figure \ref{fig:modeltree20tax} with all branch lengths
set to $0.05$.   These distributions are left-tailed, and the mean increases and the 
standard deviation decreases as $k$ increases.    Behavior for other values of $k$ is similar.
The red dots show the scores for the true $k$-splits in the tree, which are the smallest
in the data.  (There are no $6$-splits in the tree.)
The blue line segment marks the mean.  The $x$-axis is the same on all subplots, but the $y$-axis
scales differ vastly because $\binom{20}{k}$ increases as $k$ varies from $2$ to $10$.
(See online for color.)}\label{fig:splitscoredist}
\end{figure*}
\end{linenomath*}

\smallskip

\noindent {\bf Split scores distributions:} 
The histograms (Fig. \ref{fig:splitscoredist}) also shed light on the
distribution of split scores.  Notably, as $k$ increases, the mean,
which is shown in blue in Figure \ref{fig:splitscoredist}, of the
scores increases and the spread of the scores narrows.

As observed above, there are solid theoretical reasons why the size
$k$ of a $k$-split should affect the range of score values for both
true and false splits, with smaller $k$ giving rise to smaller scores.
This phenomenon can be explained in part by our algebraic-geometric
understanding of the dimension of the space of $m \times n$ matrices
of rank $4$ or less that fundamentally underlies the development of
our methods.

\smallskip

\noindent \textbf{Effect of split size:} 
In Table \ref{tab:mean_and_std}, the mean and
standard deviations are displayed for all $\dsp \binom{20}{k}$ $k$-splits for
a single data set, emphasizing in numerical terms the effect of $k$ on the
distributions. While one might naively expect scores for all size splits
to be comparable, there is a clear
pattern of larger scores when the split is closer to being
``balanced'' with equal numbers of taxa on each side of the edge.
This prompts a caution to any practitioner:  scores for different
split sizes should not be compared.

\begin{table}[h]
\begin{center}
Mean and standard deviation of $k$-split scores

\medskip
\begin{tabular}{ccc}
Split size & mean & std\\
\hline\hline
~~2 & 0.2487 & 0.0398 \\
~~3 & 0.2997 & 0.0261 \\
~~4 & 0.3276 & 0.0202 \\
~~5 & 0.3459 & 0.0167 \\
~~6 & 0.3585 & 0.0146 \\
~~7 & 0.3672 & 0.0131 \\
~~8 & 0.3730 & 0.0122 \\
~~9 & 0.3762 & 0.0117 \\
10 & 0.3773 & 0.0115\\
\end{tabular}
\end{center}
\caption{}
\label{tab:mean_and_std}
\end{table}

\smallskip

\begin{figure*}[!ht]
\centerline{{\vbox to 355pt{\vfill\hbox to 500pt{\hfill\mbox{\includegraphics[width=.9\textwidth]{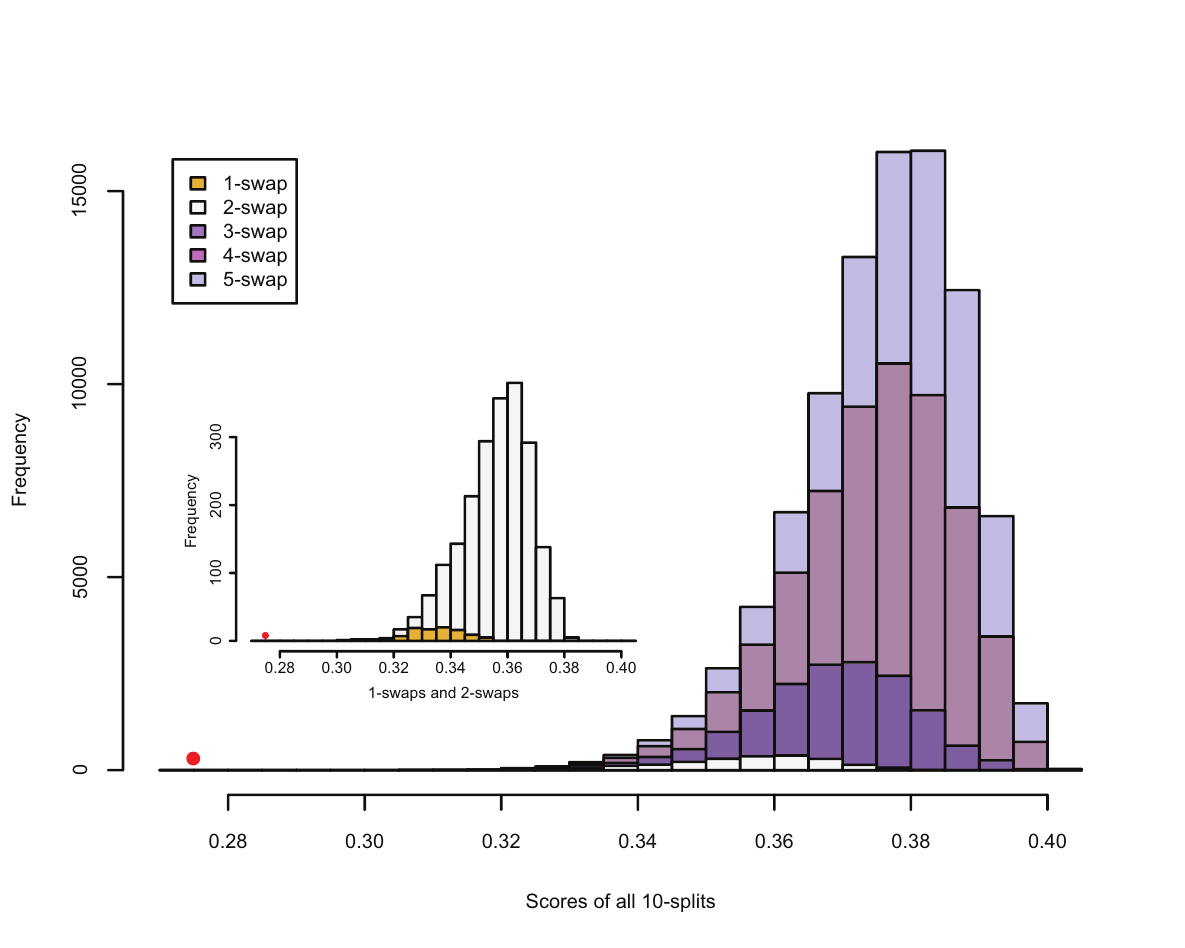}}
\hfill}\vfill}}}
\caption{Distribution of scores of all 10-splits from  data simulated on the tree of Figure \ref{fig:modeltree20tax} 
under the Jukes-Cantor model with edge lengths 0.05. The score for the one true split of size 10
 lies to the left of all others and is shown as a red dot. For false splits, colors indicate the number of individual taxa 
that are swapped between the sets of the true split. (See online for color.)}\label{fig:splitscoredist2}
\end{figure*}

\noindent \textbf{Effect of ``closeness'' to a true split} 
The histogram in Figure \ref{fig:splitscoredist2} 
displays scores for all possible 10-splits for data simulated 
from the tree in Figure \ref{fig:modeltree20tax}. 
The coloring illuminates how  scores are distributed for false
splits that differ from the true $10$-split $X_1|X_2$ 
on the tree in Figure \ref{fig:modeltree20tax} by
swapping $\ell$ taxa between the sets, for various $\ell$. 
Observations are colored according to how many taxa need to be 
swapped from a false split to produce the single true 10-split.
Note that splits that require only one or two swaps 
tend to have lower scores,
while those requiring three or four swaps tend to have higher scores. 
Splits requiring more than
four swaps to obtain the true split generally have higher scores.  Thus,
the magnitude of a score gives an indication of 
how near that split is to being a true split in the underlying tree.
%

\smallskip

\subsection*{Simulation 2 Results}

\noindent \textbf{Effect of sequence length:} Figure \ref{fig:seqlength_and_bl}a
shows split scores for two edges of the metric tree displayed in
Figure \ref{fig:modeltree20tax}, where each branch has
length $0.05$.  Since the values of scores depends on the size of
a split (cf. theoretical discussion and Simulation 1 results),
boxplots are displayed for two true splits of $T$.  The first split is
a $9$-split and corresponds to edge $e$ in Figure 1.
The second true split is a $2$-split.
The scores are computed from simulations of sequences of increasing
length, and show that scores of true splits decrease as the sequence
length grows.  This behavior is expected, assuming good model fit,
since as the sequence length grows, the empirical distribution more
closely matches the theoretical one, and the score for a true split
should approach the theoretical value 0. Shorter sequences produce
empirical distributions that are typically poorer approximations to
the asymptotics of the model.

Figure \ref{fig:sl_and_bl_swap}a shows boxplots for the difference of
false and true split scores (i.e., false-split-score minus
true-split-score) for 100 replicate data sets.  Across all sequence
lengths (500 bp, 5000 bp, 50000 bp), we see that the difference is
positive, indicating the true split score is always the smallest, even
when the false split differs little from the true split. (This held
for sequences as short as 150 bp; results not shown.)  Moreover, for
all lengths the magnitude of the score difference increases as the
false splits deviate more from the true one.  These simulations
illustrate the ability of the split score to detect the deviation of a
false split from a true split at a broad range of sequence lengths.  \
\begin{linenomath*}
\begin{figure*}[!htb]
\centerline{\vbox to 220pt{\vfill\hbox to 500pt{\hfill\mbox{\includegraphics[width=\textwidth]{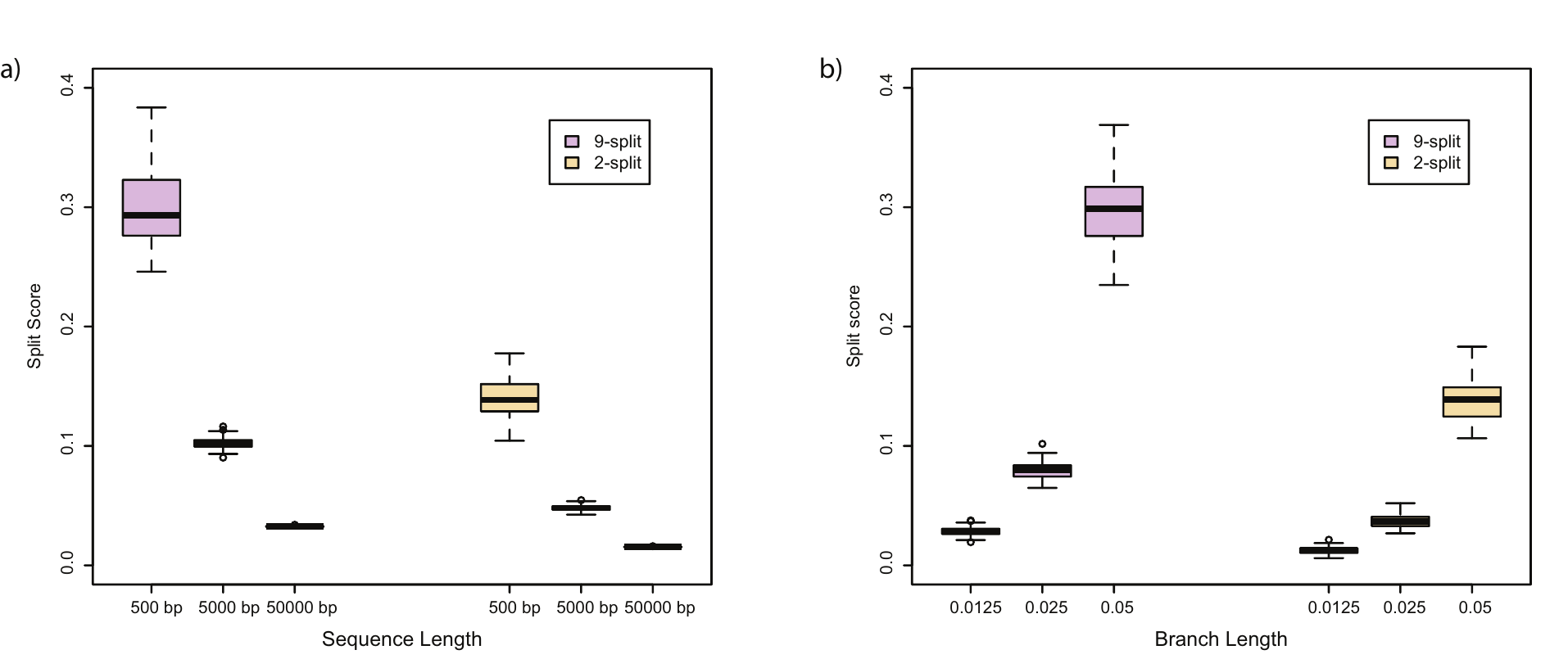}}\hfill}\vfill}}
\caption{(a) Split scores associated to two edges on tree $T$ (Fig. \ref{fig:modeltree20tax})
from sequences of length 500, 5000, 50000 bp simulated on $T$ according to the Jukes-Cantor model, 
with all edge lengths 0.05. Each boxplot represents scores for 100 data sets.
This illustrates that split scores for true splits decrease with sequence length. 
(b) The distribution of split scores for two true splits in three trees are shown.  The three trees
are topologically identical (Fig. \ref{fig:modeltree20tax}) with all branch lengths set to
$0.0125$, $0.025$, $0.05$ respectively.  As tree diameter increases, so do the score values.
This should be expected since the amount of mutation present in data scales with tree diameter,
and mutation obscures the rank $4$ signal.  (See online for color.)
}\label{fig:seqlength_and_bl}
\end{figure*}
\end{linenomath*}

\smallskip

\begin{linenomath*}
\begin{figure*}[!htb] 
\centerline{\vbox to 375pt{\vfill\hbox to 500pt{\hfill\mbox{\includegraphics[width=.85\textwidth]{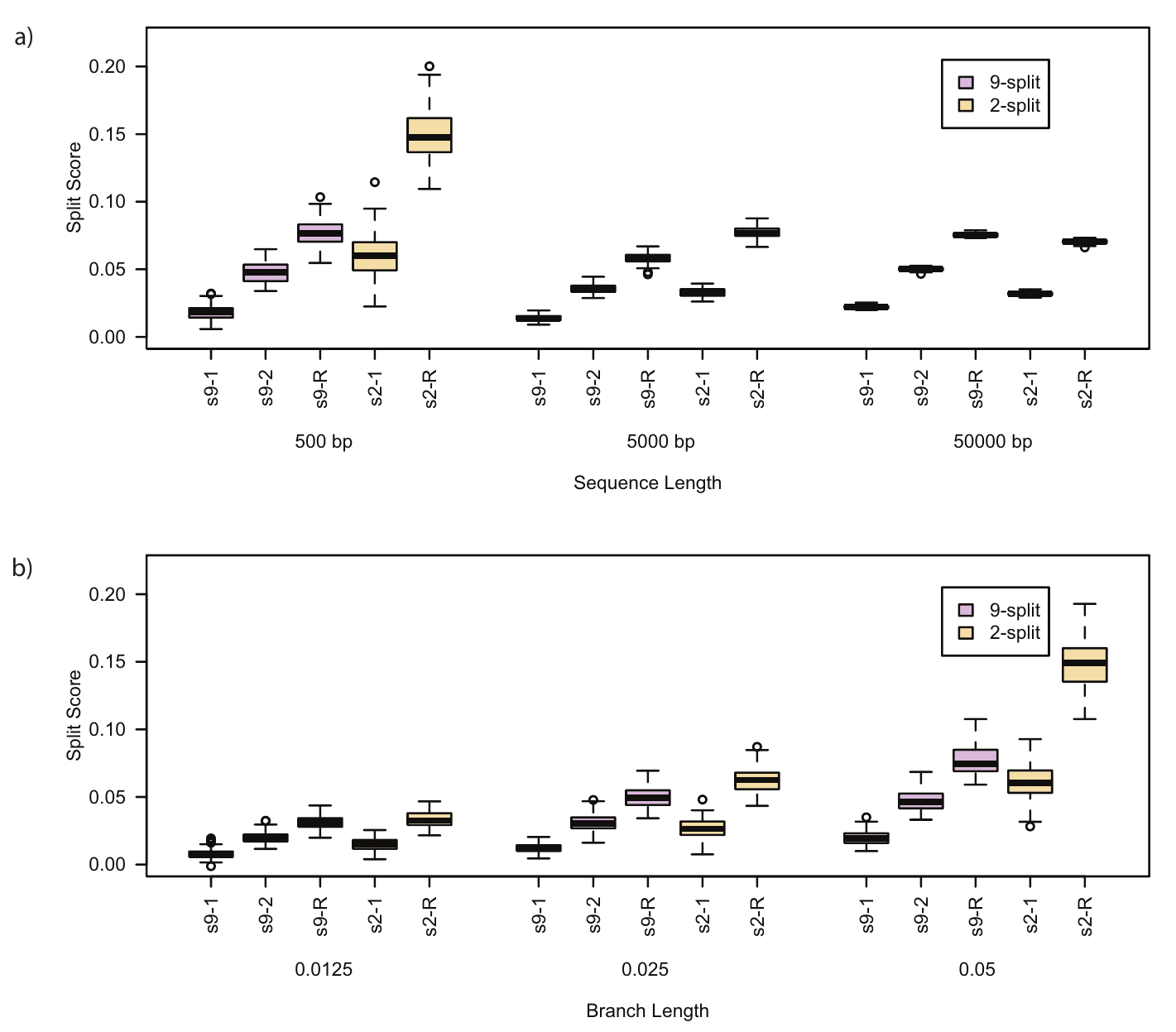}}\hfill}\vfill}}
\caption{(a) Differences (false-split-score minus true-split-score) for true 9- and 2-splits associated to two edges of tree $T$ of Fig.~\ref{fig:modeltree20tax},
from sequences of length 500, 5000, 50000 bp simulated on $T$ according to the Jukes-Cantor model, 
with all edge lengths 0.05.  Each boxplot represents differences for 100 replicates.
See Simulation Study 2 methods for choices of false and true splits.
In these simulations the differences were always positive, illustrating that split scores for true splits are the 
smallest over a range of sequence lengths.
(b) Boxplots illustrating the distribution of the score differences as in (a), but with sequence length 500 bp and
all branch lengths set to $0.0125$, $0.025$, $0.05$.  
Across this range of branch lengths, with a single exception (one simulation for s9-1 and length 0.0125), the true split score
was always the smallest.  (See online for color.)
}\label{fig:sl_and_bl_swap}
\end{figure*}
\end{linenomath*}

\noindent \textbf{Effect of tree diameter:} Figure \ref{fig:seqlength_and_bl}b
shows the distribution of split scores for two  true splits in three trees with identical
topologies.  All the trees have the topology shown in Figure \ref{fig:modeltree20tax},
but branch lengths have been scaled by a fixed factor, increasing the
tree diameter.  As might be anticipated, the scores increase with tree diameter
since longer branch lengths produce more site substitutions, diluting the signal
that a flattening matrix is close to rank $4$. 
(With extremely long branch lengths, 
as saturation is reached, all scores will drop as the matrix rank goes to $1$.)

Each cluster of boxplots in Figure \ref{fig:sl_and_bl_swap}b shows
score differences (false-split-score minus true-split-score) for five
false splits, with branch lengths varied across simulations (0.0125,
0.025, 0.05).  With a single exception (for the false 9-split closest
to true, s9-1, and shortest branch lengths) the difference was
positive, indicating that the score of the true split was the smallest.
This indicates that the split score can distinguish true splits from
false splits over a range of branch lengths.

\smallskip

\begin{figure*}[!htb]
\centerline{{\vbox to 400pt{\vfill\hbox to 500pt{\hfill\mbox{\includegraphics[width=.88\textwidth]{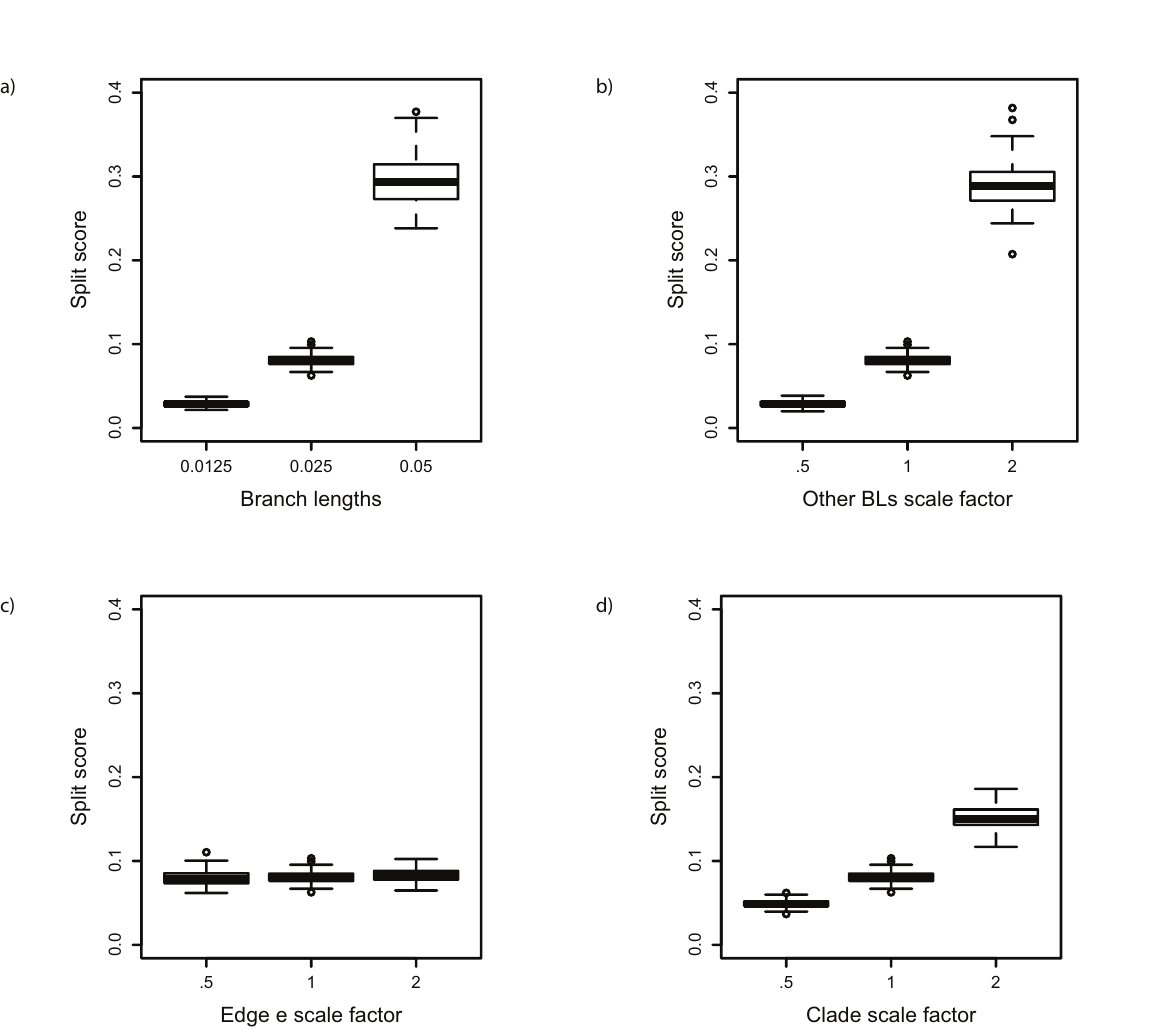}}\hfill}\vfill}}}
\caption{Split scores associated to a single edge $e$ in the tree from Figure \ref{fig:modeltree20tax}
computed from 500 bp sequences simulated according to the Jukes-Cantor model.   The
base tree in Figure  \ref{fig:modeltree20tax} has all branch lengths equal to $0.025$.
The single edge $e$ is the one that induces the split  that bipartitions the taxa into groups
of size $9$ and $11$,
$\{t_1, t_{14}, t_{11}, t_{17}, t_{3}, t_{13}, t_{16},  t_{19}, t_{18} \} \mid 
\{ t_{2}, t_{20}, t_{6}, t_{8}, t_{9}, t_{5}, t_{12}, t_{4}, t_{15}, t_{7}, t_{10}\}$.
In each panel, scores for edge $e$ in the three trees are shown; these trees are topologically identical
but certain branch lengths have been modified in each tree:  a) All branch lengths in the trees are equal, though
of differing magnitude, changing the tree diameter; 
b) all branch lengths EXCEPT that corresponding to edge $e$ are scaled; c) ONLY the 
branch length that corresponds to edge $e$ is scaled; and d) all branch lengths in the 
clade corresponding to taxa 
$\{t_1, t_{14}, t_{11}, t_{17}, t_{3}, t_{13}, t_{16},  t_{19}, t_{18} \}$ are scaled. In each subfigure, boxplots show 
distributions of scores for 100 simulated data sets. }\label{fig:depth}
\end{figure*}

\noindent \textbf{Effect of metric depth in tree:} Figure
\ref{fig:depth} shows split scores for a single edge $e$ of three
trees, all with the topology shown in Figure \ref{fig:modeltree20tax},
but with different branch lengths.  The metric structure of the trees
differs by rescaling either all or a subset of the edges
(Fig. \ref{fig:depth}a, b, d), or by scaling only edge $e$
(Fig. \ref{fig:depth}c).  In Figures \ref{fig:depth}a, b, d as the
scaling factor is increased, the metric depth of the split in the tree
(i.e, the average distance of the split from the leaves)
increases.  These simulations show that the split score for a true
split increases with metric depth.  This behavior is not surprising,
and is consistent with the tree diameter scaling results of Figure
\ref{fig:seqlength_and_bl}b as the deeper an edge lies in a tree, the
more obscured evidence for it may be by base substitutions nearer the
leaves of the tree.  In Figure \ref{fig:depth}c, only the edge $e$ is
scaled, while the other branches all have length $0.025$.  Since the
metric depth of $e$ is held fixed, only small variation in the value
of the score for the split induced by edge $e$ is observed.
%

\subsection*{Simulation 3 Results}

\noindent \textbf{Detecting changes in the evolutionary process:} 
\begin{linenomath*}
\begin{figure*}[!ht]
\centerline{\vbox to 200pt{\vfill\hbox to 800pt{\hfill\mbox{\includegraphics[width=.95\textwidth]{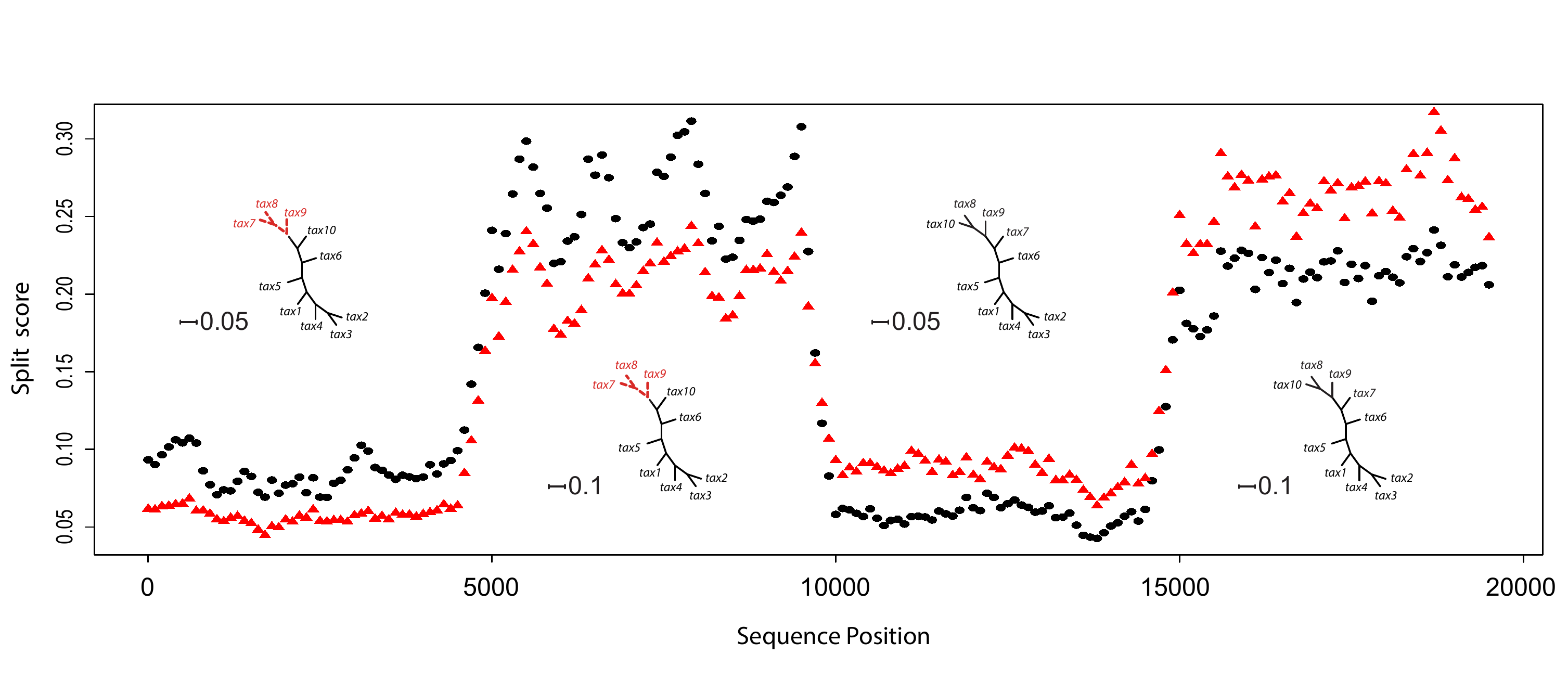}}\hfill}\vfill}}
\caption{The values of the scores for the split \{7-9\} $\mid X \smallsetminus$ \{7-9\}  are
shown as red triangles, and for the split \{8-10\} $\mid X \smallsetminus$ \{8-10\}
as black dots.  The scores were computed by scanning along the 20,000 bp simulated genome
and computing scores for the \{7-9\} grouping and \{8-10\} grouping for segments of length 500 bp
at a slide size of 100 bp.  For each 100 bp --- $1, 101, 201, \dots$ --- on the $x$-axis,
both scores for the 500 bp window beginning at that site in the genome are plotted.
The score detects both the true split in the tree used to generate the sequence data, and
the shift in branch lengths.  (See online for color.)
}\label{fig:SW}
\end{figure*}
\end{linenomath*}
The plot displayed in Figure \ref{fig:SW} shows that our score detects
multiple changes in the evolutionary process along the genome.  In
particular, we see for the first 10,000 bp or so that the true split
\{7-9\} in the tree used to generate this portion of the genome has a
lower score than the false split (i.e., the red triangle is lower than the
black dot).  After 10,000 bp, the black dot has the smaller value;
here the score captures that the sequence data for 10,000 - 20,000 bp
was generated on trees with the \{8-10\} split.

Interestingly, our score detects not only changes in tree topology but
also changes in numerical parameters of the evolutionary model.  The
genome sections corresponding to the first and third quarters of the
sequence data (1-5,000 bp; 10,001-15,000 bp) were generated with
branch lengths set to $0.05$, while the second and fourth quarters of
the data (5,001-10,000 bp; 15,001-20,000 bp) were generated on trees
with branch lengths $0.1$.  The scores for both splits are `small' for
the trees with small tree diameter, and `large' for the trees with
large branch lengths, consistent with the results in Figure
\ref{fig:seqlength_and_bl}b.

Because our simulation and computations of scores used a sliding
window of length 500 bp, but the sequence data was generated with an
abrupt change in evolutionary model at sites 5,001, 10,001, and
15,001, we see a gentle rise (or fall) in the values of the scores
around these transition points in our plot.  This reflects that a
window of size 500 bp will contain a number (400, 300, 200, 100) of
sites generated from one tree, and a number (100, 200, 300, 400) of
sites generated from a tree that differs in either topology or branch
lengths, when the sliding window overlaps a transition site.  This
highlights that the selection of a window size and slide size are
important parameters for analyses of this sort.  Significantly, with
properly chosen parameters, Figure \ref{fig:SW} supports the
hypothesis that our score can detect rough boundaries that signify
shifts in the underlying evolutionary process.

\smallskip

\subsection*{Applications to Empirical Data}

\begin{figure*}[!htb]
\centerline{\includegraphics[width=.9\textwidth]{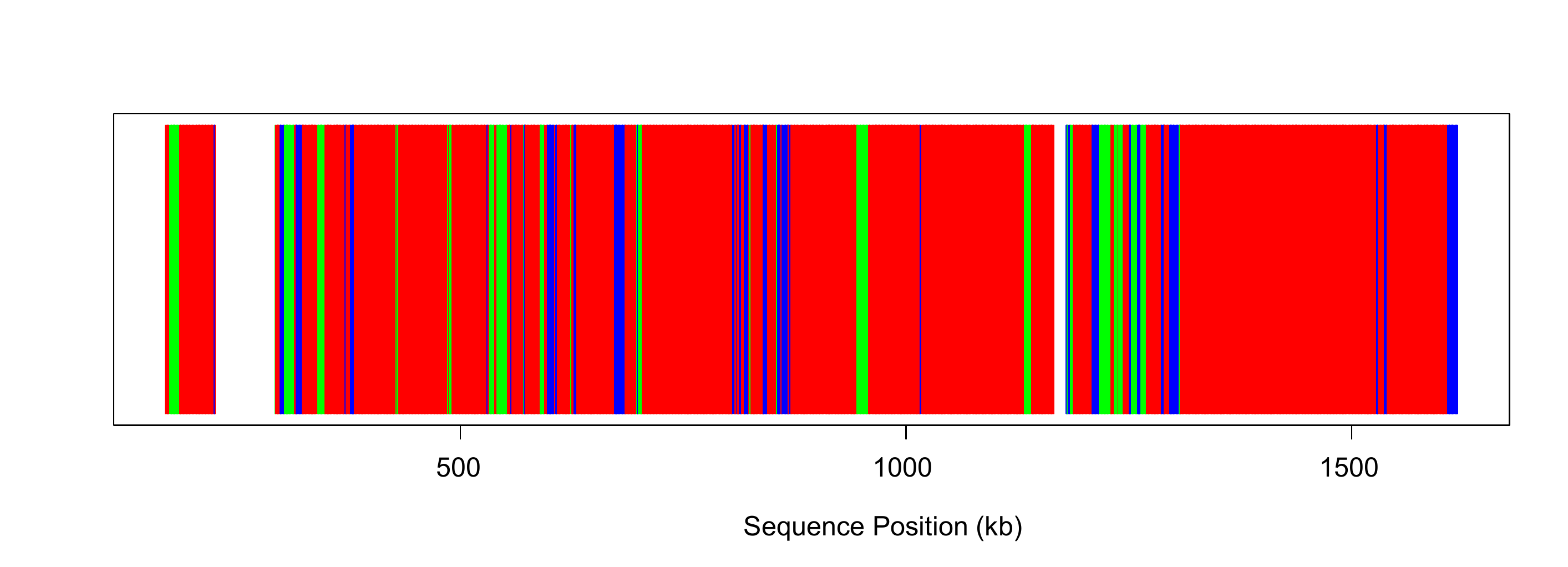} \hskip .5cm \ }
\caption{Application of our sliding window method to the primate data set.  At each location along the chromosome, a red vertical line is drawn if the tree with the lowest score has the human-chimp clade, a green vertical line is drawn if the tree with the lowest score has the chimp-gorilla clade, and a blue vertical line is drawn if the tree with the lowest score has the human-gorilla clade. White indicates locations with excessive gaps.  This data set exemplifies the expectation under the coalescent model, in that the gene tree that agrees with the species tree is expected to be more frequently observed in a set of three taxa, with the two alternative topologies occurring with equal frequency.
(See online for color.) }\label{fig:primate.empex}
\end{figure*}

\noindent {\bf Primate Data} -- Figure \ref{fig:primate.empex} shows
the results of the analysis of chromosome 7 for the four primate
taxa. At each location along chromosome 7, a red vertical line
indicates that the lowest score for that window corresponds to the
split that contains the human-chimp clade, a green vertical line
indicates that the lowest score corresponds to the split that contains
the chimp-gorilla clade, and a blue vertical line indicates that the
lowest score corresponds to the split that contains the human-gorilla
clade.  In this example, we expect to see variation along the
chromosome as predicted by the coalescent model.  In particular,
because the human-chimp clade is well-established as the true
phylogenetic relationship, we expect the majority of the locations
along the chromosome to show this relationship with the two other
relationships arising with the same, lower frequency, as is easily
observed from the figure.

\smallskip

\begin{figure*}[!htb]
\centerline{\includegraphics[width=.9\textwidth]{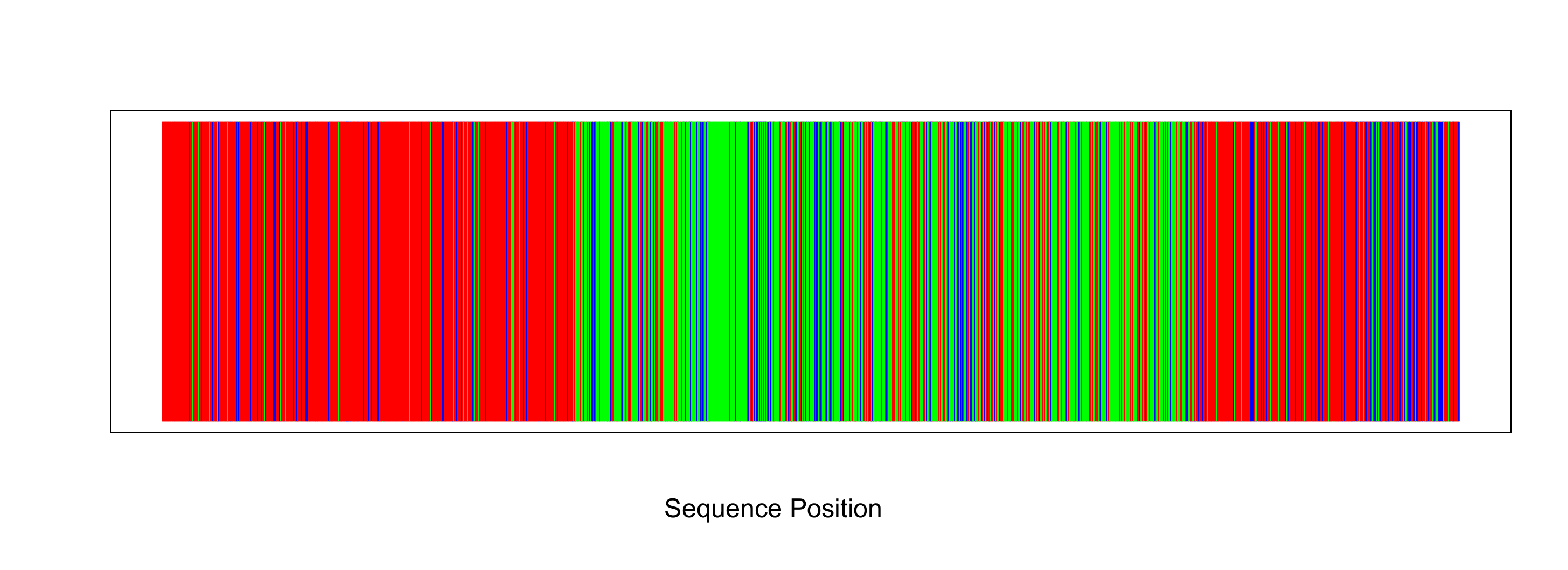}}
\caption{Application of our sliding window method to the mosquito data set.  At each location along the chromosome, 
a red vertical line is drawn  if the tree with the lowest score has the {\itshape An. gambiae-An. coluzzii} clade, a blue vertical line 
is drawn if the tree with the lowest score has the {\itshape An. gambiae-An. arabiensis} clade, and a green vertical line is 
drawn if the tree with the lowest score has the 
{\itshape An. coluzzii-An. arabiensis} clade. Some species are known to have experienced a chromosomal inversion in a 
portion of this region of chromosome 2L, and the method easily picks out the location of the inversion as a shift in 
the phylogeny.  (See online for color.) }\label{fig:mosquito.empex}
\end{figure*}

\noindent {\bf Mosquito Data} --  Figure  \ref{fig:mosquito.empex} shows the results of analyzing 
chromosome 2L for the four mosquito species. The plot is organized as described 
for the primate data, with the red vertical lines corresponding to the split that contains 
the {\itshape An. gambiae}-{\itshape An. coluzzii} clade, the blue vertical lines 
corresponding to 
the split that contains the {\itshape An. gambiae}-{\itshape An. arabiensis} clade, 
and the green vertical lines corresponding to the 
split that contains the {\itshape An. coluzzii}-{\itshape An. arabiensis} clade. The 
most striking feature of the graph is the center region, in which the dominant 
phylogeny is that containing the {\itshape An. coluzzii}-{\itshape An. arabiensis} clade. 
This finding agrees with the results of \citet{fontaineetal2015} 
(see their Figure 2 and Figure 5), for which the majority of 
chromosome 2L shows  {\itshape An. gambiae} and {\itshape An. coluzzii} to be sister taxa (indicated
by the red vertical lines in Figure  \ref{fig:mosquito.empex}), but a
chromosomal inversion in a portion of chromosome 2L in {\itshape An. arabiensis} leads to a 
closer relationship with the sample from {\itshape An. coluzzii} (indicated by the green vertical lines
in Figure  \ref{fig:mosquito.empex}), which shares this inversion, over
that portion of the chromosome.

This sliding window analysis was performed on a data set of size over 37.5 million bp,
and sparse matrix flattenings were constructed for 37,556 windows, each of length 10,000 bp 
and of which 37,547 had more than 500 gapless sites so that scores were computed.  
The computation time for a single pairing, say the {\itshape An. gambiae}-{\itshape An. coluzzii} clade,
was 7.28 minutes on a MacBook Pro 3.1 GHz processor with 16 GB of memory.

\smallskip

\begin{figure*}[!htb]
\centerline{\includegraphics[width=.9\textwidth]{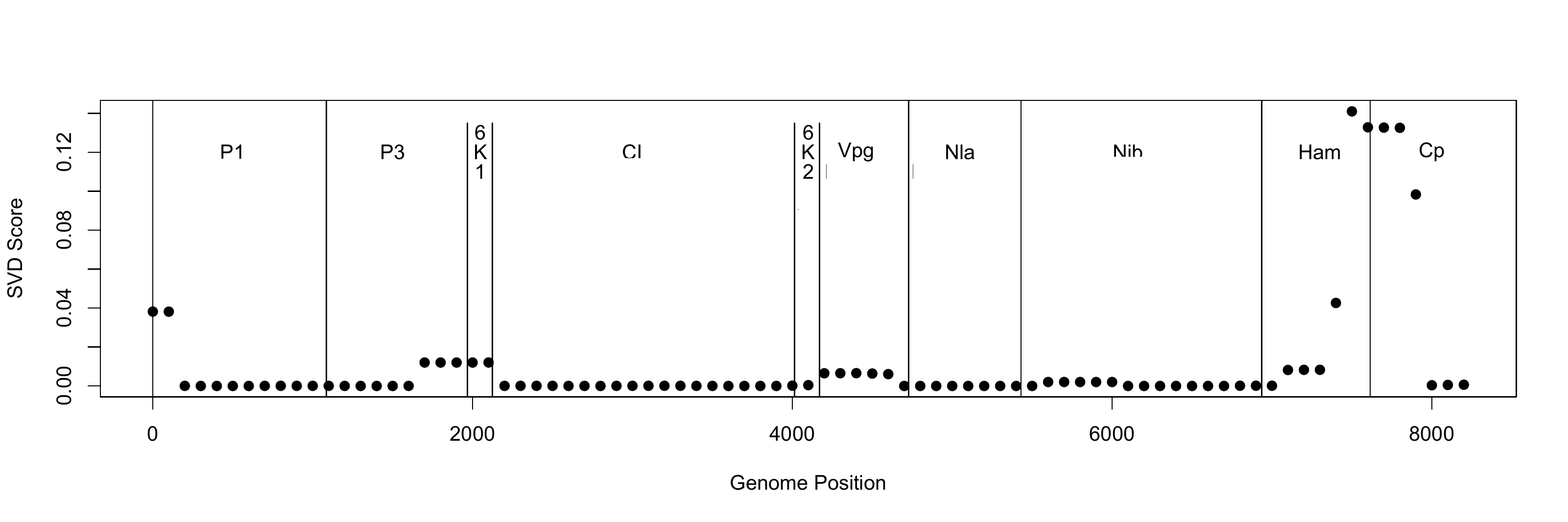}}
\caption{Application our sliding window method to the CBSV data set. We evaluate support for the split that partitions the 
sequences into CBSV vs.~UCBSV at locations across the complete viral genome.  Gene boundaries are given by 
black vertical lines, with gene labels presented at the top of the graph. The value of the likelihood ratio statistic for testing 
for a difference in the ratio of synonymous vs.~nonsynonymous substitutions is also given, with a `*' indicating those genes 
for which the likelihood ratio test is significant at the 5\% level.}\label{fig:cbsv.empex}
\end{figure*}

\noindent {\bf CBSV Data} -- Figure  \ref{fig:cbsv.empex} shows the results of analyzing the 29 viral genomes, 
with black vertical lines delimiting boundaries between genes.  Points on the plot are the score
for the split that partitions the CBSV sequences from the UCBSV sequences.   It is easy to see that shifts in the scores correspond largely 
to boundaries between genes, indicating potential shifts in the corresponding evolutionary
processes governing mutation rates in CBSV vs. UCBSV, in agreement with the results of 
\citet{alicaietal2016}.  This supports the 
results of the simulations shown in Figure \ref{fig:SW}, in which the score was shown to vary based on 
shifts in either the topology or the evolutionary model parameters.

Also shown on the plot are the likelihood ratio statistics for each gene for the test of differing 
synonymous/nonsynonymous substitution rates for the CBSV vs.~UCBSV clades \cite{yang1997}. 
Statistics that are significant at 5\% level are indicated with a `*'.  The split score is correlated with 
significance of the likelihood ratio test, in the sense that lower scores are associated 
with significant results for many genes.  
This result thus indicates variation in the evolutionary process along the genome, and hints 
that changes in mutation rates may be driving this variation.

\section*{Discussion}

We have presented the split score as a means of quantifying the
strength of the biological signal for specific splits on a
phylogenetic tree. We demonstrate that while the score is affected by
the amount of data (i.e., number of sites), the lengths of the
branches in the tree, and the size of the split under consideration,
the score can accurately indicate which splits are most strongly
supported by a given data set.  Importantly, the split score can be
computed extremely rapidly, because it requires only counting of site
patterns in order to construct the flattening matrix and computation
of singular values from the flattening matrix.  Thus, the split score
is well-equipped to handle the genome-scale data sets that are being
generated today.  We view the split score as a valuable tool for
exploratory analysis of genome-scale data sets of arbitrary size.

We have presented three empirical examples that demonstrate a
practical application of our method. These involve evaluating the
split score at various locations along a contiguous alignment in a
``sliding window'' analysis. The three examples demonstrate the
varying types of biological phenomena that can be detected by an
analysis such as this.  The primate data show the pattern expected in
a typical species tree analysis, where only incomplete lineage sorting
causes variation across a chromosome. The mosquito data show that
variation in the underlying evolutionary process (in this case, a
chromosomal inversion) can also be detected by the method, though it
is clear that the method indicates only variation in the process and
does not indicate the cause of such variation.  Finally, the CBSV data
set demonstrates application of the method to more than four taxa (in
this case, 29 taxa) and shows that the method can detect shifts in the
underlying evolutionary process even when the topology remains fixed.

An important characteristic of all three empirical data sets is that
they represent genome-scale data: the primate data set consists of an
alignment of $\sim$1.9 million bp for 4 taxa, the mosquito data set
consists of $\sim$37.5 million bp for four taxa, and the CBSV data set
consists of approximately 9,000 bp for 29 taxa. In all cases, the
entire sliding window analysis can be carried out within minutes on a
standard desktop machine, providing a huge computational advantage
over other phylogenetic tools that seek to extract similar
information.  We provide freely available software that requires only
a PHYLIP-formatted input file and a list of splits to be evaluated to
allow others to use this exploratory tool.

\subsection*{Detecting true splits in a tree} 

While it would be highly desirable to understand the dependency of the
distribution of true split scores from data under a fixed model of
base substitution, even with an assumption of a specific tree topology
and edge lengths, this question is a complex one that is probably not
addressable theoretically.  (One could, of course, perform a
parametric bootstrap for an approximation.)  When the tree is unknown,
theoretical analysis seems even more difficult. While work has been
done on the distributions of singular values for certain types of
random matrices, the true split flattening matrices arising from
distributions along a tree, with their multinomial entries, are not
covered by these results.

One approach for interpreting a split score $S(X_1|X_2)$ for a split
that is not known to be true or false is to compare it to the
distribution of scores $S(X_1'|X_2'),$ $|X_i'|=|X_i|$ for all splits
of the same size computed from the same data.  The false splits among
these should produce larger scores than the true ones (of which there
may be several.) This is borne out by Figure \ref{fig:splitscoredist}
which shows several distributions of split scores from a simulated
data set on the tree in Figure \ref{fig:modeltree20tax}. When true
splits of a given size exist, they are markedly below the rest of the
distribution. When no true splits exist, there are no such outliers.
In preliminary trials, we found after some naive normalizations, such
as computing z-scores, that true split scores are markedly smaller
than the closest false split scores.  Though we were not able to
provide any probabilistic bounds on the difference between normalized
true and false split scores, such ideas hold promise and need further
statistical development.  One step in this direction is provided by
\citet{gaitherkubatko2016}, who develop formal statistical hypothesis
tests for splits of four taxa under the coalescent model.

One should be careful in interpreting plots like that of Figure
\ref{fig:splitscoredist}. They allow one only to compare whether a
given split is more supported than alternative splits \emph{of the
  same size}. A tree may have no splits of a given size ({\it e.g.}
there are no $6$-splits in the model tree), and so the lowest score
should not be interpreted as indicating a split that should be on the
tree, but only that it is supported more than other splits.

When the number of taxa is not too large (e.g., $<26$) one can quickly
compute full distributions of all split scores for a fixed split
size. For a large number of taxa, however, obtaining the exact
distribution may be computationally prohibitive. In such a case, a
method of approximating the distribution is suggested by Figure
\ref{fig:splitscoredist2}.  Suppose we wish to assess whether a
particular $k$-split is supported or not.  We compute the split score
distribution for all splits differing from it by swapping $1\le \ell
\le L$ taxa, for some chosen $L$.  We then compute split scores for a
large number of random splits of the same size as the one to be
assessed (possibly discarding those that arise from swaps already
considered when $k$ is small {\it i.e.} sample without replacement for
small splits).  We then use an appropriate weighted combination of the
random and small-swap distribution as an approximation to the full
one.  If the given split is a low outlier in comparison to this
approximation, we view it as supported.
The full small-swap distribution helps us obtain an accurate
approximation at the lower end of the distribution, since if the given
split is true, these tend to give lower scores, but would not be
well-represented among random splits.

\subsection*{Extensions}

As discussed in the Results section, the split score has the potential
for use in phylogenetic inference, although we have not pursued that
possibility here.  \citet{eriksson2005} presented an algorithm based
on a non-normalized variant of the split score for inferring a gene
tree from the alignment for a single gene, but that method has
significant weaknesses.  Since the method fails to take into
consideration the differing dimensions of the varieties of all
possible $k$-splits and compares splits of differing size, it is
strongly biased towards returning more balanced trees, replete with
cherries and small clades, since scores for $2$-splits and $3$-splits
are generally the smallest.  In part because of that bias, more recent
uses of the SVD of flattenings for tree inference have so far focused
on quartet-based inference
\cite{casanellasfernandezsanchez2007,casanellasfernandezsanchez2015,
  chifmankubatko2015}, so that split size issues do not arise.

This said, the split score holds promise for use beyond the specific
goals of the three empirical studies presented here.  New ideas and
development of alternative ways to use its fast computation and
ability to detect splits or near-splits in data sets are still needed.
As one example, the score of potential splits to be evaluated in a
heuristic procedure that searches over tree space could be rapidly
computed, and the most promising ``direction'' for the search as
indicated by the split score could then be rigorously evaluated using
a more standard model-based criterion, such as Maximum Likelihood.
Indeed, this was the idea that first motivated this work.  Overall, it
is clear that the split score contains information that can be used to
differentiate true from false splits, and its rapid computation time
makes it a promising tool for phylogenetic inference.

A natural extension of the results presented here applies to certain
models even more general than the GM model.  For instance, under a
$2$- or $3$-class mixture of GM models on the same tree, the rank of
matrix flattenings corresponding to edges in the true tree $T$ are of
$8$ and $12$ respectively \cite{ARidtree}.  While the GM model is
already parameter-rich --- and consequently unfamiliar to many ---
particular submodels of these GM-mixure models, such as GTR+I, are in
wide use.  The covarion model \cite{TuffleySteel1998} also has
well-understood flattening ranks \cite{ARidtree}, so a similar split
score can be used for it.  To facilitate use with these models, the
software {\tt SplitSup} was designed with an optional parameter to set
the rank used for an analysis to values other than its default of 4.

The findings presented here are likely to apply to related work, as
well. For example, \citet{chifmankubatko2015} used a similar split
score to indicate support for splits of 4 taxa under the coalescent
model.  The primary difference between their version of the split
score and that presented here is that the rank of the flattening
matrix corresponding to a true split is 10, rather than 4, in order to
accommodate gene tree variability due to the coalescent process.
Their score would then indicate support for splits in the species
tree, rather than the gene tree as considered here.  The score
computed under the coalescent model is likely to behave in similar
ways with regard to properties such as effect of sequence length,
effect of changes in the substitution model, etc., to the score shown
here.  As this model underlies the SVDQuartets method for
coalescent-based species tree inference that is implemented in PAUP*
\cite{swofford2016} and being increasingly used for species-level
phylogenetic inference, it is important to understand behavior of the
split score in detail.

\medskip

In conclusion, phylogenetic invariants and, more generally, methods
based on relationships in observed site pattern frequencies, are
increasingly relevant to genome-scale phylogenetic inference.  These
offer two advantages for data collected at the genome scale. First,
these methods perform better as more data become available, because
each site pattern probability is better approximated by its observed
frequency as the sample size increases. Second, computations
underlying tools such as the split score can scale extremely well as
the number of nucleotides and/or taxa increases, as all that is
required is counting of observed site patterns and application of
well-developed efficient methods for matrix calculations.  We thus
recommend continued study of methods based on phylogenetic invariants
and the algebraic properties of site pattern probabilities arising
from phylogenetic models, as these show promise for new
computationally-efficient approaches to genome-scale phylogenetic
inference.

\section*{F{\sc UNDING}}
E.~Allman and J.~Rhodes were supported in part by the National Institutes of Health 
grant R01 GM117590, awarded under the  Joint DMS/NIGMS Initiative to Support Research 
at the Interface of the Biological and Mathematical Sciences. L.~Kubatko's work was funded
in part by National Science Foundation grant 11-06706.

\section*{A{\sc CKNOWLEDGMENTS}}
We thank Dr.~Dennis Pearl for his instrumental role in starting this
collaboration.  He brought the authors together following a conference
at the Mathematical Biosciences Institute, and suggested many fruitful
directions of investigation for developing and using splits scores.
We are grateful to him for his intellectual contributions to this
project.

We also thank Dr.~Laura Boykin and her collaborators for sharing the CBSV data set.

\bibliographystyle{abbrvnat}
\setcitestyle{authoryear,open={((},close={))}}
\bibliography{SplitScores}

\end{document}